\documentclass[iop]{emulateapj}
 \usepackage{amsmath, amsthm, amssymb,amsfonts}
\usepackage{graphicx}
\usepackage{dcolumn}
\usepackage{bm}
\usepackage{color}
\usepackage[abs]{overpic}

\begin{document}

\title{Nonlinear firehose relaxation and constant-B field fluctuations}

\author{Anna Tenerani}
\affiliation{Department of Earth, Planetary, and Space Sciences, UCLA, Los Angeles, CA, 90095}
  \email{annatenerani@epss.ucla.edu}
\author{Marco Velli}%
 \affiliation{Department of Earth, Planetary, and Space Sciences, UCLA, Los Angeles, CA, 90095}

\date{\today}

\begin{abstract}

The nonlinear evolution of Alfv\'enic fluctuations in the firehose unstable regime is investigated numerically and theoretically for an anisotropic plasma described by the one-fluid double adiabatic equations. We revisit the traditional theory of the instability and examine the nonlinear saturation mechanism, showing that it corresponds to evolution towards states that minimize an appropriate energy functional. We demonstrate that such states correspond to broadband magnetic and velocity field fluctuations with an overall constant magnitude of the magnetic field. These nonlinear states provide a basin of attraction for the long-term nonlinear evolution of the instability, a self-organization process that may play a role in maintaining the constant-$B$ Alfv\'enic states seen in the solar wind in the  high-$\beta$ regime.

\end{abstract}

\maketitle

\section{Introduction} 
Alfv\'enic fluctuations are incompressible, non-dispersive perturbations of plasma velocity (${\bf U}$) and magnetic field (${\bf B}$) such that ${\bf \delta U}\propto\mp {\bf \delta B}$~\citep{alfven}. They are ubiquitous in magnetized plasmas and the study of their stability and nonlinear dynamics provides the building blocks to understand turbulence in many collisionless and weakly collisional astrophysical environments~\citep{scheko_2006}, solar coronal heating~\citep{rappazzo_2018} and solar wind acceleration~\citep{verdini_2010}.

Since the beginning of the space age, the solar wind has remained the most explored collisionless, turbulent natural plasma. Fast solar wind streams are typically pervaded by large amplitude Alfv\'enic fluctuations propagating away from the sun that display a well defined power-law spectrum in frequency, the latter characterized by a low and high-frequency part with spectral indices close to $-1$ and $-5/3$, respectively~\citep{bruno}. At the same time particle distribution functions display many non-thermal features~\citep{marsch}, like pressure anisotropy, that may provide the free energy  for the onset of kinetic instabilities possibly coexisting with the evolving turbulent spectrum~\citep{matteini_2013,hellinger_2015}. 
 
The large scale expansion of the solar wind naturally drives plasma distribution functions towards the threshold of the oblique  or parallel proton firehose instability~\citep{hellinger_2015}, a similar driving to threshold also occurring, e.g. via shear flow,  in other astrophysical environments like accretion disks~\citep{kunz}, making the study of such an instability of particular interest. The firehose instability arises in high-$\beta$ (thermal to magnetic pressure ratio) plasmas when the pressure anisotropy ($p_\parallel>p_\bot$,  $p_\parallel$ and $p_\bot$ being the pressure parallel and perpendicular to the magnetic field) is large enough to remove the restoring force due to magnetic tension~\citep{parker_1958}, leading to an increase of magnetic energy until marginal stability is reached nonlinearly. Observations around 1~AU~\citep{kasper_2002, hellinger_2006} seem to support the idea that the firehose may play a role in controlling the anisotropy in the high-$\beta$ wind. How the transition to the unstable regime affects the evolution of large amplitude Alfv\'enic fluctuations has however remained unexplored until the present Letter.

Particle dynamics and magnetic field variations are strongly coupled in collisionless (or weakly collisional) plasmas via the adiabatic invariants $(B^2p_\parallel)/\rho^3$ and $p_\bot/(B\rho)$~\citep{cgl} ($\rho$ being the plasma density and $B$ the total magnetic field  magnitude) that are conserved to lowest order (in length and time scales with respect to the ionic typical scales). Such a coupling introduces a nonlinear feedback on fluctuations mediated by the bulk response of particles to variations of $B$  that dramatically depends on the  polarization of the fluctuations. 

In this regard, \citet{squire} studied finite amplitude effects of a linearly polarized Alfv\'en wave in collisionless and weakly-collisional high-$\beta$ plasmas.  They found that, above a threshold amplitude, the Alfv\'en  wave induces a firehose instability that would ultimately lead to the ``interruption'' of the wave itself. In this process the wave evolves into a sequence of spatial discontinuities (square profiles) that minimize the gradients of $B^2$, a tendency previously observed also in hybrid numerical simulations~\citep{vasquez_1998}.

On the other hand, one of the peculiarity of Alfv\'enic fluctuations in the solar wind is that the magnitude of the total magnetic field remains nearly constant, i.e.,  $\delta B/B\ll1$. This implies that the tip of the magnetic field vector is bound to rotate on a sphere of radius $B$~\citep{tsuru}. Such a condition  corresponds to spherical polarization, that in the simplest case of  plane fluctuations reduces to circular (for parallel propagation) or arc polarization (for oblique propagation).  A similar geometrical behavior is found also in the velocity vector when doppler shifted to the wave frame, pointing to the fact that in such a frame the particle energy is conserved~\citep{matteini_2015}. Although it is known that nonlinear  Alfv\'enic fluctuations with constant total pressure represent an exact solution of the MHD equations~\citep{barnes_1974}, how such a nonlinear state can be accessed dynamically and maintained remains to be understood. 

In this Letter we explore a nonlinear relaxation process that may play a role in preserving constant-$B$ states at large~$\beta$ values. We investigate both numerically and theoretically the evolution of Alfv\'enic fluctuations subject to the firehose instability. By relaxing the constraint of linear polarization we show that broadband, constant-$B$ nonlinear states provide a basin of attraction of the long-term evolution of the firehose instability.

\section{Setup of the problem}
In this work  we adopt the one-fluid double adiabatic model first introduced by~\citet{cgl} (CGL model) which, in the limit of long wavelength and small frequencies with respect to the typical proton spatial and temporal scale, provides  a  good starting point to investigate pressure anisotropy effects~\citep{hunana_2017}. 

We  start from the full set of the CGL equations, where an explicit diffusion due to kinematic viscosity ($\nu$) and magnetic diffusivity ($\eta$) is introduced for numerical stability, ${\bf \hat b}={\bf B}/B$ and $\Delta p=p_\bot-p_\parallel$:
\begin{equation}
\frac{\partial\rho}{\partial t}+\boldsymbol\nabla\cdot{\bf (\rho U)}= 0
\label{CGL_eq0}  
\end{equation}
\begin{equation}
\begin{split}
&\frac{\partial {\bf U}}{\partial t}+{\bf U}\cdot\boldsymbol{\nabla}{\bf U}=-\frac{1}{\rho}\boldsymbol{\nabla}\left(p_\bot+ \frac{B^2}{8\pi}\right)+\\
&\frac{1}{4\pi \rho}{\bf B}\cdot \boldsymbol{\nabla}{\bf B}+\frac{1}{\rho}\boldsymbol\nabla\cdot\left( {\bf \hat b\hat b}\Delta p \right)+\nu\nabla^2{\bf U},
\end{split}
\end{equation}
\begin{equation}
\frac{\partial {\bf B}}{\partial t}=\boldsymbol{\nabla}\times({\bf U\times B})+\eta\nabla^2{\bf B},
\end{equation}
\begin{equation}
\frac{d }{d t}\left( \frac{p_\parallel B^2}{\rho^3} \right)=0,\quad \frac{d }{d t}\left( \frac{p_\bot }{\rho B} \right)=0. 
\label{CGL_eq1}  
\end{equation}

A spatial dependence along the $z$ coordinate is assumed and the background magnetic field ${\bf B}_0$ is taken in the $(z,y)$ plane, forming an angle $\theta_0$ with the $z$-axis. In an ideal plasma ($\eta=\nu=0$), finite amplitude, one-dimensional (plane)  fluctuations transverse to the propagation direction are described by the following equations:  
\begin{equation}
\begin{split}
\frac{\partial {\bf U_{\bot}}}{\partial t}=&\frac{B_{0}\cos\theta_0}{4\pi\rho}\frac{\partial}{\partial z}\left[\left(1+\frac{4\pi}{B^2}\Delta p\right){\bf B_{\bot}}\right],
\end{split}
\label{cgl1}
\end{equation}

\begin{equation}
\begin{split}
\frac{\partial {\bf B_{\bot}}}{\partial t}=B_{0}\cos\theta_0\frac{\partial}{\partial z}{\bf U_{\bot}}.  
\label{cgl2}
\end{split}
\end{equation}

In the specific case in which $B$ is spatially uniform there are no compressible effects and the pressure term $\Delta p$ is a function of time only. Therefore, one can  easily  integrate eqs.~(\ref{CGL_eq1}) to obtain the following expression,
\begin{equation}
\Delta p(t)=\left[ p_{0\bot}\sqrt{\frac{B^2(t)}{ B^2(0)}}- p_{0\parallel}\frac{B^2(0)}{ B^2(t)}   \right].
\label{cgl4}
\end{equation}

Equations~(\ref{cgl1})--(\ref{cgl2}) together with eq.~(\ref{cgl4}) generalize the MHD equations of  Alfv\'enic  fluctuations with arbitrary amplitude and in the presence of pressure anisotropy, and they exactly describe the dynamics of circularly or arc-polarized nonlinear states~\citep{anna1}. Note  that eqs.~(\ref{cgl1})--(\ref{cgl2}) are valid also when  $B$  is not uniform in space. In this case however one has to take into account the coupling with compressible modes and the full complement of the double adiabatic and continuity equations would have to be considered. In a high-$\beta$ plasma coupling with compressible modes is however negligible, and we will see {\it a posteriori} that  eqs.~(\ref{cgl1})--(\ref{cgl4}) can be used in the more general case of non-constant-$B$ fluctuations to study their evolution.

For the forthcoming discussions it is useful to introduce the normalized magnetic field and   total magnetic field magnitude, ${\bf \hat B_{\bot}}={\bf B_{\bot}}/B_{0}$, $\hat B=B/B_0$, the Alfv\'en speed $V_a=B_{0}/\sqrt{4\pi\rho}$ and  
\begin{equation}
\tilde V_a^2=V_a^2+\frac{1}{\rho}\frac{\Delta p}{\hat B^2}. 
\label{margi}
\end{equation}

Next,  combine eqs.~(\ref{cgl1})--(\ref{cgl2}) in the following second order partial differential equation:
\begin{equation}
\begin{split}
\frac{\partial^2}{\partial t^2}{\bf \hat B_{\bot}}=(\cos\theta_0)^2\frac{\partial^2}{\partial z^2}\left( \tilde V_a^2{\bf \hat B_\bot}\right).
\end{split}
\label{cgl3}
\end{equation}

Equation~(\ref{cgl3}) describes nonlinear propagating fluctuations if $\tilde V_a^2>0$ that might be unstable to parametric decay~\citep{anna1}. Here we focus on the opposite case in which $\tilde V_a^2<0$ at $t=0$. This condition corresponds to the well-known firehose regime, whose threshold, given  by the condition $\tilde V_a^2=0$, is now extended to include finite amplitude effects. In the following, we first investigate the nonlinear evolution of fluctuations in such a regime by integrating  the set of eqs.~(\ref{CGL_eq0})--(\ref{CGL_eq1}) numerically starting from an initial non constant-$B$ fluctuation; next, we will use eq.~(\ref{cgl3}) to discuss the results on theoretical grounds.

 \begin{figure}[tb]
 \centering{
\includegraphics[width=0.4\textwidth]{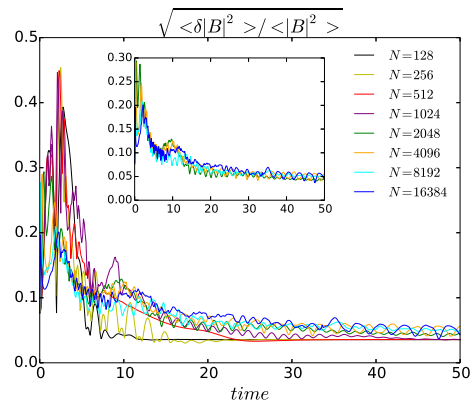}
}
\caption{Nonlinear relaxation of an Alfv\'enic fluctuation ($\theta_0=0$). Normalized root-mean-square amplitude vs. time for each simulation; high-resolution simulations are shown in the inset plot.  }.
\label{parallel}
\end{figure}

\begin{figure}[tb]
\centering{
\includegraphics[width=0.44\textwidth]{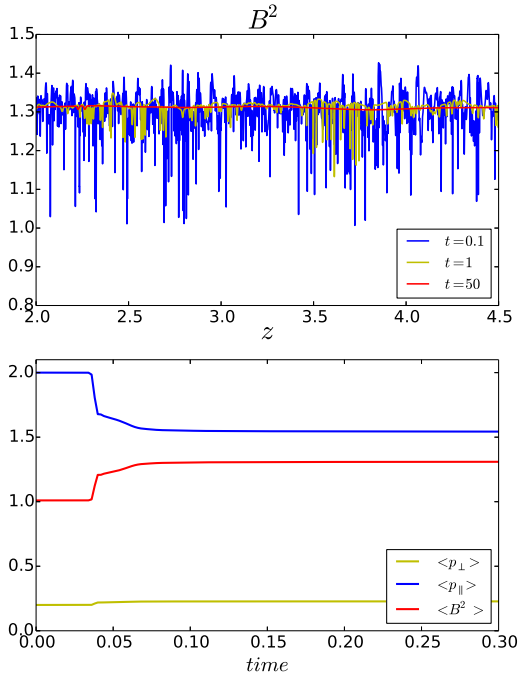}
\caption{ Results from the highest resolution simulation ($\theta_0=0$, $N=16384$). Upper panel: plot of the total magnetic field magnitude $B^2(z)$  in the subdomain $z=2-4.5$ at three different times. Lower panel: temporal evolution of the average $B^2$, $p_\parallel$ and $p_\bot$  during the linear stage and saturation.}
}
\label{macro}
\end{figure}

\section{Numerical results}
\label{sim}
 We  consider a case study in parallel propagation ($\theta_0=0$) and initialize the simulation with uniform density $\rho$ and relatively large amplitude magnetic and velocity field fluctuations of the form
\begin{equation}
{\hat B_x}= A\cos(k_0z),\quad {\hat B_y=A\sin(k_0 z+\delta\varphi)}, 
\label{ic1}
\end{equation}
\begin{equation}
{ U_x}=-{\hat B_x}\sqrt{\mathcal C},\quad  {U_y}=-{\hat B_y}\sqrt{\mathcal C}, 
\label{ic2}
\end{equation}
where $\mathcal C=|<\tilde V_a^2(0)>|$ and brackets $\left<\cdot\right>$ denote the spatial average. The amplitude is fixed to $A=0.1$, the initial anisotropy $\xi\equiv p_{0\bot}/p_{0\parallel}=0.1$ and $\beta_\parallel\equiv8\pi p_{0\parallel}/B_0^2=4$, so that $\tilde V_a^2(0)<0$ (speeds are normalized to $V_a$). We chose  $k_0=3$, and we perturbed the circular polarization by  imposing  $\delta\varphi=1$.  The length of the box is fixed to $L=2\pi$  and we performed eight simulations with increasing resolution by changing the number of mesh points from $N=128$ to $N=16384$, and decreasing the magnetic diffusivity  ($\eta$) and kinematic viscosity ($\nu$) accordingly. The Prandtl number is one ($\nu=\eta$) and the diffusion coefficients range from $\eta=0.027$ to $\eta=0.0002$. 

 In Fig.~\ref{parallel}  we plot  the time evolution of the normalized root-mean-square amplitude of the  magnetic field magnitude fluctuation ($[\left<(B^2-\left<B^2\right>)^2\right>^{1/2}/\left<B^2\right>]^{1/2}\equiv \left<\delta  B\right>$) for each simulation.  $\left<\delta B\right>$ has a peak at the early stage, due to the growth of broad-band incoherent  fluctuations which are induced by nonlinear mode coupling (including wave steepening), as will be explained in the next section. After the peak,  the system relaxes towards the same final nonlinear state where $\left<\delta B\right>\ll1$ over a timescale which becomes independent of the dissipation for increasing  resolution (decreasing dissipation).   We estimated that the relative fluctuations of the magnetic field magnitude reaches the $5\%$ level around $t=30$ for $N=2048-16384$. This can be seen by inspection of the inset plot, where we show $\left<\delta B\right>$ for the higher resolution simulations.

 \begin{figure*}[t]
 \centering{
\includegraphics[width=0.328\textwidth]{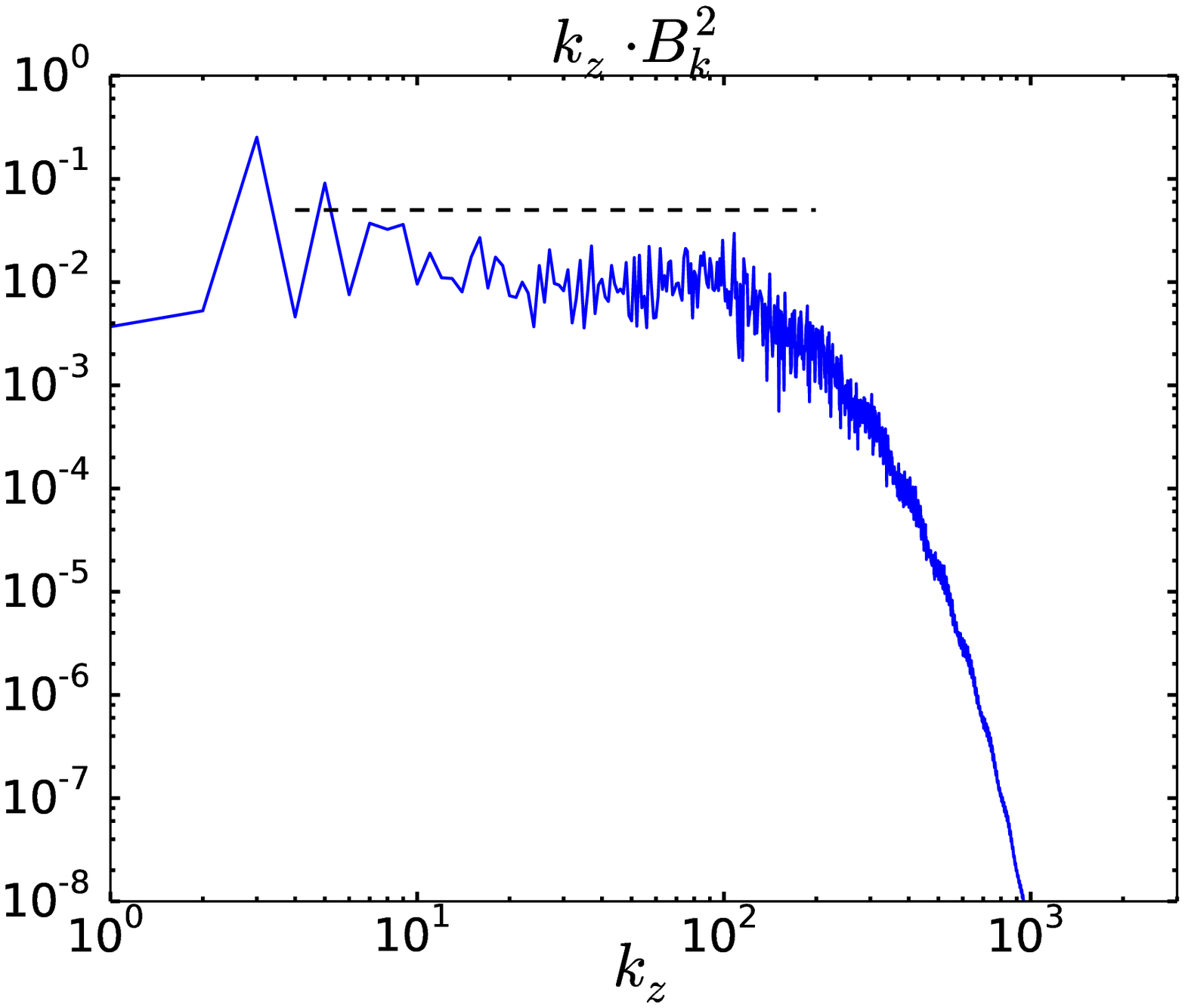}
\includegraphics[width=0.328\textwidth]{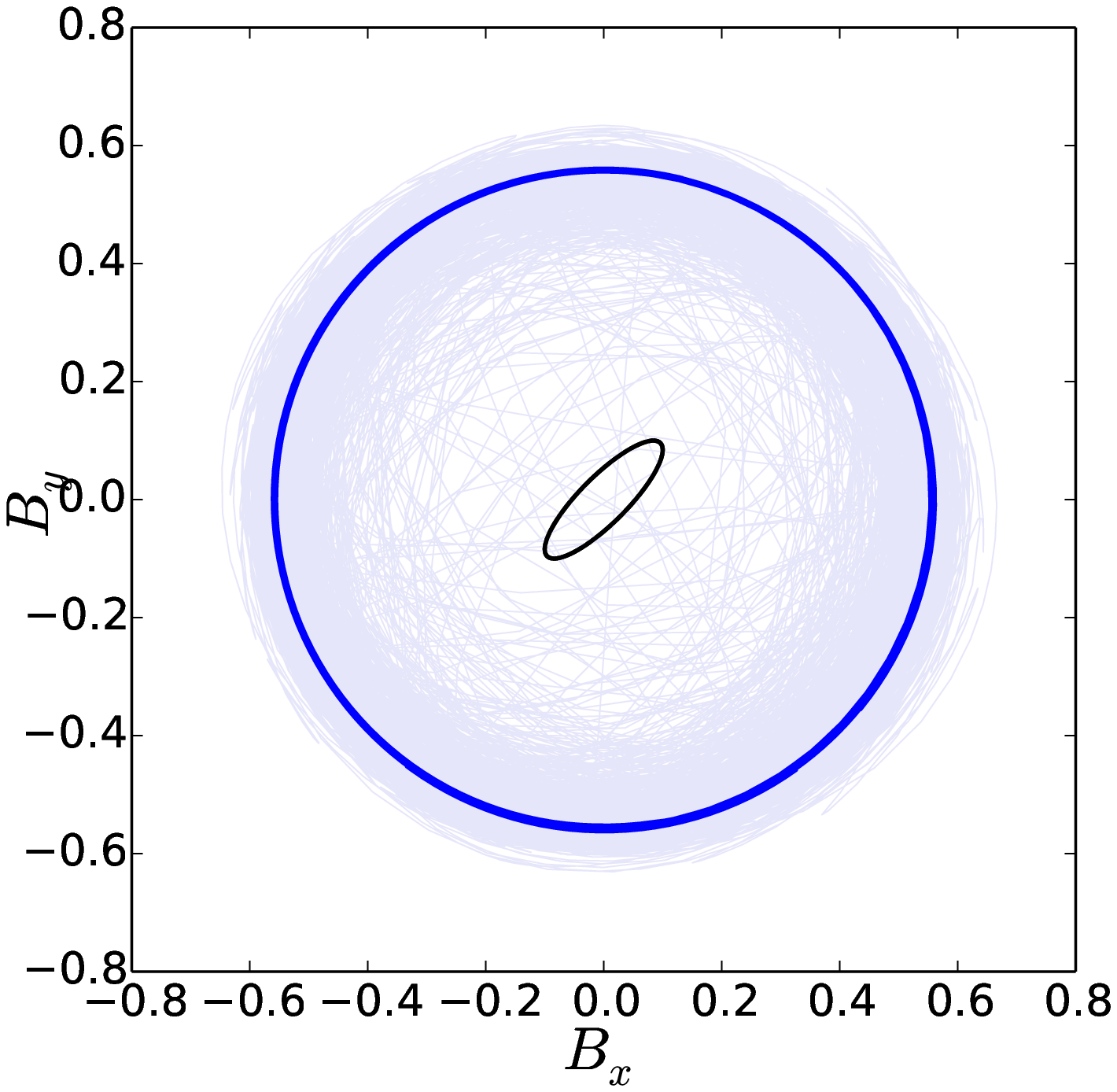}
\includegraphics[width=0.328\textwidth]{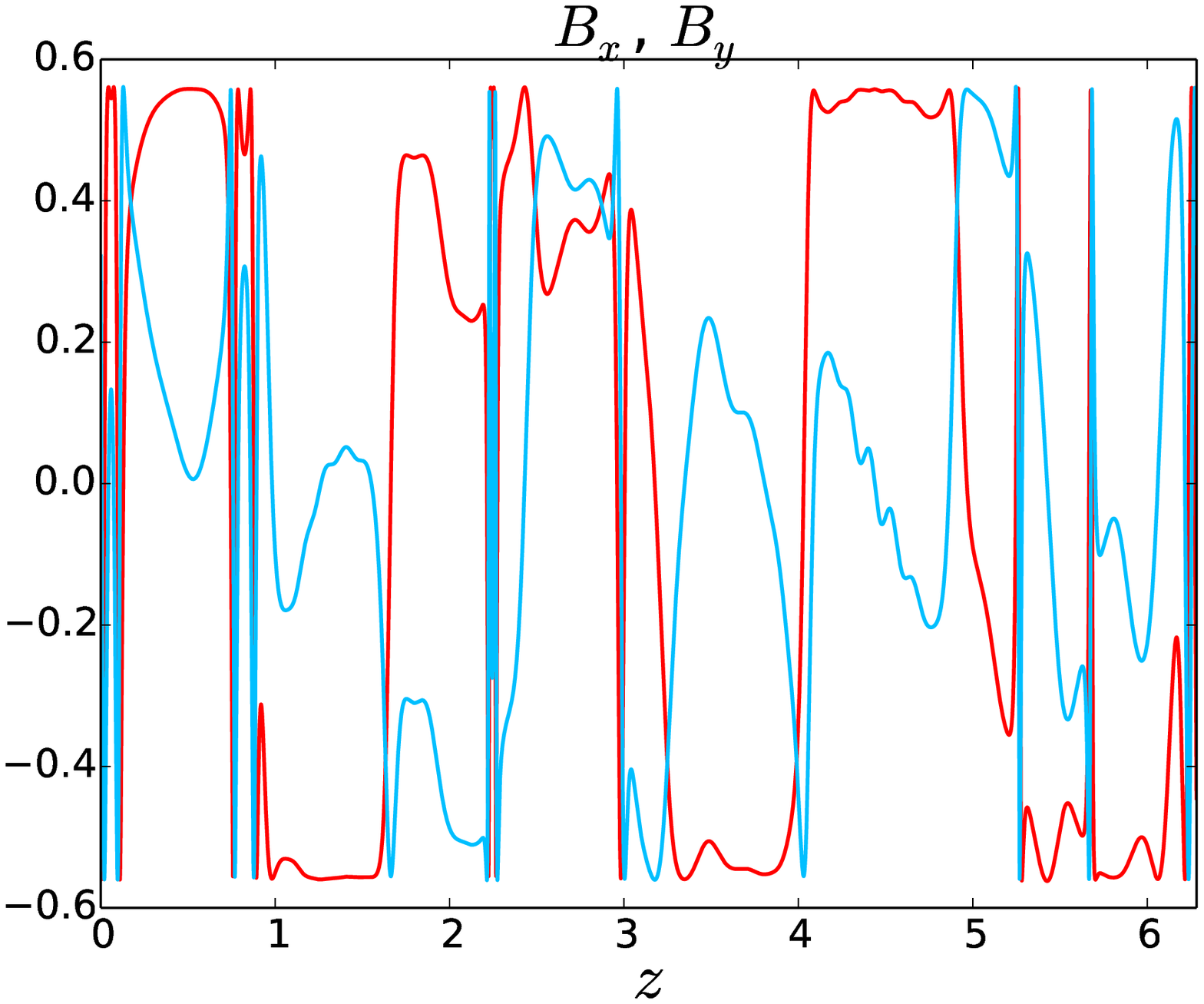}
}
\caption{Final state from the highest resolution simulation ($\theta_0=0$, $N=16384$). Left panel: compensated magnetic energy spectrum mediated between $t=40$ and $t=50$. Middle panel:  hodogram of the magnetic field at $t=0$ (black), $t=0.1$ (light blue) and $t=50$ (blue). Right panel:  magnetic field components $B_x(z)$ (light blue) and $B_y(z)$ (red)  at $t=50$.}
\label{para16}
\end{figure*}

Fig.~\ref{macro} and Fig.~\ref{para16} display numerical results corresponding to the simulation with the highest resolution ($N=16384$ mesh points and $\eta=0.0002$), which is used as a reference. Fig.~\ref{macro}, upper panel, shows the spatial structure of $B^2(z)$ in the subdomain  $z=2-4.5$ at three different times, showing that  $B^2$ gradually tends, from a highly variable profile reached at the end of the linear stage (cfr. $B^2(z)$ at $t=0.1$), towards a smooth profile (cfr. $B^2(z)$ at  $t=50$). The lower panel of Fig.~\ref{macro} shows the evolution of $\left< B^2\right>$ and of the average parallel and perpendicular pressure $\left< p_\parallel\right>$ and $\left< p_\bot\right>$: at saturation (about $t\simeq0.1$), the average magnetic field has reached its asymptotic value $\left<B^2\right>\simeq1.31$; at the same time, consistently with the increase of magnetic energy and the conservation of the two adiabatic invariants (initially $\left< B^2(0)\right>=1.01$), the initial anisotropy is reduced,  with the parallel pressure decreasing from its initial (homogeneous) value $p_\parallel(0)=2$ to $\left< p_\parallel\right>\simeq1.54$, and the perpendicular pressure slightly increasing from $p_\bot(0)=0.2$ to $\left< p_\parallel\right>\simeq0.227$.
 
 In Fig.~\ref{para16} we show in more detail the final state. The left panel displays the compensated magnetic energy spectrum  $k_z B_k^2$ mediated in time in the interval $t=40-50$, showing the generation of a developed spectrum of fluctuations whose spectral index is close to $-1$.  The middle panel represents the hodogram of the magnetic field at three different times: at $t=0$ (black); at $t=0.1$ (light blue) during the early nonlinear stage when the magnetic fluctuations are uncorrelated;  at $t=50$ (blue) when the system has reached its final state characterized by circular polarization,  implying that the fluctuating components of the magnetic field are properly shifted in phase so as to lead to a constant $B^2$ profile. In the right panel the final magnetic field waveform is shown at $t=50$: light-blue and red colors correspond to the $B_x$ and $B_y$ components, each of which  has evolved towards large amplitude fluctuations with embedded  rotational discontinuities.  We found a similar nonlinear evolution for different values of $k_0$, for a broadband initial fluctuation, and starting from initial incoherent noise. 

\begin{figure}[tb]
\includegraphics[width=0.45\textwidth]{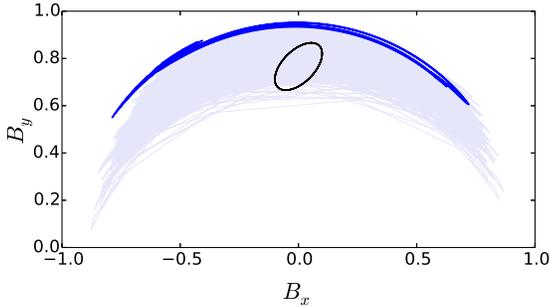}
\caption{ Magnetic field hodogram for  $\theta_0=50^\circ$ at $t=0$ (black), $t=0.2$ (light blue), and  $t=90$ (blue).}
\label{oblique}
\end{figure}

 We performed also a simulation by imposing the same initial condition given in eqs.~(\ref{ic1})--(\ref{ic2}) but choosing  $\theta_0=50^\circ$ to inspect whether the relaxation towards constant-$B$ field survives for oblique fluctuations (with $\eta=0.0004$, $N=8192$ mesh points). Although the evolution occurs on longer timescales, it is remarkable that the system  evolves again towards a constant-$B$ state with the same spectral properties. In this case the constant-$B$ condition corresponds to arc-polarized fluctuations. This is  shown in Fig.~\ref{oblique}, that displays the magnetic field hodogram at three different times: the black color corresponds to the initial condition ($t=0$), the light blue color corresponds to the firehose destabilization of small scale fluctuations ($t=0.2$),  and  blue color corresponds to the asymptotic state ($t=90$).

\section{Discussion}
Our simulations confirm that the constant-$B$ state represents a basin of attraction for the long term evolution of fluctuations that are in the firehose regime. Regardless of the initial condition, the system naturally evolves towards a stationary, broadband state whose magnetic field magnitude is almost constant in space, resulting in a specific phase-correlation between the components of ${\bf B _\bot}$ that leads to a circularly or arc-polarized magnetic field. In a previous work we have investigated theoretically the behavior of  a monochromatic and constant-$B$ initial state in the firehose regime~\citep{anna1}. We showed that, in that simplified case, eq.~(\ref{cgl3})  can be integrated exactly and that the solution for the amplitude $B(t)$ can be described by analogy with the motion of a particle moving in a conservative field of force. The amplitude is  homogeneous in space (by construction) and oscillating in time between a minimum  and a maximum value corresponding to the two turning points of the potential energy: any form of dissipation would cause the amplitude to decay towards the minimum of the potential, leading to a final stationary state of minimum energy. Although this is a simplified case, we expect a similar behavior also in the more general case of an initial non-monochromatic and non constant-$B$ field. If $B$ is not constant however there is an additional effect due to its gradient. This can be readily seen by inspecting the functional form of $\tilde V_a^2(B^2)$ (eq.~\ref{margi}): regions corresponding to maxima of $B^2$ will grow (or propagate, when $\tilde V_a^2>0$) slower than those corresponding to minima of $B^2$. Such a nonlinear dispersion naturally leads to wave steepening and in the specific case of the firehose regime it tends to flatten the profile of $B^2$. 

Let us consider  the dynamical equation for nonlinear transverse fluctuations,  eq.~(\ref{cgl3}), complemented by eqs.~(\ref{cgl4})--(\ref{margi}). We approximate $B^2(0)\approx\left< B^2(0)\right>$ and separate $B^2$ into an average and fluctuating part, $B^2=\left<B^2\right>+(B^2-\left<B^2\right>)\equiv \left<B^2\right>+\delta B^2$. After developing $\tilde V_a^2$ to first order in $\delta B^2$ and transforming into Fourier space, from eq.~(\ref{cgl3}) we get the following set of coupled nonlinear equations for each  Fourier mode ${\bf B}_{k}$,
\begin{equation}
\begin{split}
\ddot  {\bf B}_{k}=&-\frac{d\phi_{k}}{d{\bf B}_k} + (\cos\theta_0)^2(\tilde V_a^2)^{\prime} \sum_{p\neq k, q}{\bf B}_p({\bf B}_{p-k-q}\cdot{\bf B}_q), 
\label{moto}
\end{split}
\end{equation}

where for the sake of simplicity the {\it hat} has been dropped, the dot denotes the time derivative, $(\tilde V_a^2)^\prime=d\tilde V_a^2/dB^2$  evaluated on the mean field, and the potential $\phi_k$, by writing the normalized mean square  field $\left<B^2\right>=1+\sum_{q}|{\bf B}_{q}|^2$ and $\omega_a=kV_a\cos\theta_0 $, is given by 
\begin{equation}
\begin{split}
 \phi_k=&\frac{\omega_a^2}{2}  \left[|{\bf B}_{k}|^2+ \frac{\beta_\parallel}{2}\left(2\xi\frac{\sqrt{\left<B^2\right>}}{\sqrt{ \left<B^2(0)\right>}}+\frac{\left<B^2(0)\right>}{\left< B^2\right>}\right)\right],
\end{split}
\label{cgl6}
\end{equation}

The first term on the right-hand-side of eq.~(\ref{moto}) introduces a mean-field coupling among the existing Fourier modes and it drives amplitude oscillations  in a way similar to the monochromatic case discussed above, where the Fourier modes can be seen alike a system of oscillators coupled via the potential $\phi_k$. In this case $\phi_k$ is a paraboloid that for $\xi<1$ and $\beta_\parallel>1$ has a maximum at the origin and a minimum on the hyper-surface defined by $\tilde V_a^2(\left<B^2\right>)=0$. The last term  on the right-hand-side instead arises from the inhomogeneity of $B^2$. This nonlinearity is at the core of the long-term nonlinear evolution of firehose fluctuations since it couples a given mode with wavenumber $k$ with  higher wavenumber modes, conveying energy across scales all the way down to the dissipative ones and leading to wave steepening. The cascade process introduces an effective dissipation that allows the system of oscillators to ``fall" towards the minimum energy state corresponding to $\tilde V_a^2(\left<B^2\right>)=0$, that generalizes the otherwise known marginal stability condition. Marginality predicts the average square magnetic field magnitude at saturation, that in our simulation is $\left< B^2\right>\simeq1.31$ (see Fig.~\ref{macro}, lower panel). While approaching marginal stability, the same nonlinearity ultimately leads towards a constant-$B$ state. This process can be better understood by writing the corresponding equation in real space. At marginal stability, and by retaining only the dominant contribution from the gradient of $\delta B^2$, eq.~(\ref{moto})  can be approximated by $\partial^2_t{\bf  B_{\bot}}\approx (\cos\theta_0)^2(\tilde V_a^2)^\prime{\bf B_\bot}\partial^2_z B_\bot^2$. As can be verified for non-propagating fluctuations, according to this equation magnetic field in regions close to minima of $B^2$ tend to increase while those close to maxima tend to decrease. Notice that \citep{squire} found a similar description of the nonlinear evolution of a linearly polarized Alfv\'en wave subject to the (self-induced) firehose instability. However,  a linearly polarized fluctuation is constrained to have $B_\bot^2=0$ at minima. As a consequence, the nonlinearity   tends to form a squared magnetic field waveform, leading to flat profiles of $B^2$ spaced out by deep magnetic holes in correspondence of  the  field reversals. Instead, in the more general case with more than one fluctuating component maxima and minima of $B^2$ are shifted with respect to crests and null points of the field ${\bf B}_\bot$. This allows the fluctuating components to undergo a phase shift that leads to a smooth, constant profile of the total $B^2$.  

In the regime studied here an initial Alfv\'enic fluctuation evolves into a non-propagating fluctuation, and therefore the final state does not show the  correlation $\delta{\bf B}\propto\delta {\bf U}$. This is to be expected since if $\tilde V_a^2<0$ in the whole dominium then eq.~(\ref{cgl3}) is a non-propagative (elliptic) differential equation. Finite amplitude effects however may lead to intermediate cases in which the plasma is only partially unstable, with $\tilde V_a^2<0$ in localized regions around minima of $B^2$. In this case it is reasonable to expect that the evolution would display both a tendency to constant-$B$ states while keeping a high degree of Alfv\'enic correlation (cross helicity). The final cross helicity and polarization may depend on initial conditions. This is a study that deserves further development in future work. 

 The results presented here rely on a one-fluid description of the plasma. Kinetic effects are known to lower the threshold for the onset of  the firehose  with respect to the CGL prediction. In that regime, the firehose leads to the growth of fluctuations at the small kinetic scales, while the large ``MHD" scales remain unaffected~\citep{hunana_2017}. We argue that, in this case, firehose fluctuations may grow on top of large amplitude Alfv\'enic fluctuations and lead nonlinearly to the relaxation of the initial anisotropy. A fluid behavior should be recovered in the presence of a statistically relevant number of particles resonating with the unstable spectrum. In those cases our results should remain valid on the large scales. However, the impact of dispersion and other kinetic effects on our results (and, vice-versa, the effect of large amplitude Alfv\'enic fluctuations on kinetic instabilities) remains an open question that is deferred to  later work.

\section{Summary}We have investigated the nonlinear evolution of Alfv\'enic fluctuations subject to the firehose instability in the CGL limit: the instability naturally leads via a nonlinear self-organization process to a relaxed state characterized by a spectrum of fluctuations  whose spectral index is close to $-1$ and with magnetic field components displaying defined phase shifts  corresponding to constant magnetic field.  In the  solar wind, such a nonlinear relaxation due to the firehose may provide a way to preserve constant-$B$ states as Alfv\'enic fluctuations propagate outwards towards the high-$\beta$ regions.

\begin{acknowledgements}
The authors would like to thank P. Hellinger for useful discussions. This research was supported by the NASA Parker Solar Probe Observatory Scientist grant NNX15AF34G and used the Extreme Science and Engineering Discovery Environment (XSEDE) Comet at the San Diego Supercomputer Center through allocation TG-AST160007. XSEDE is supported by National Science Foundation grant number ACI-1548562. 
\end{acknowledgements}

\end{document}